\documentclass[conference]{IEEEtran}
\usepackage[latin9]{inputenc}
\usepackage{array}
\usepackage{url}
\usepackage{multirow}
\usepackage{amsmath}
\usepackage{amssymb}
\usepackage{graphicx}
\usepackage{esint}

\makeatletter

\DeclareTextSymbolDefault{\textquotedbl}{T1}
\providecommand{\tabularnewline}{\\}

\IEEEoverridecommandlockouts
\usepackage{cite}
\usepackage{amsfonts}
\usepackage{algorithmic}
\usepackage{textcomp}
\usepackage{xcolor}
\def\BibTeX{{\rm B\kern-.05em{\sc i\kern-.025em b}\kern-.08em
    T\kern-.1667em\lower.7ex\hbox{E}\kern-.125emX}}

\makeatother

\begin{document}
\title{A physics-informed generative model for passive
radio-frequency sensing\thanks{Funded by the European Union. Views and opinions expressed are however those of the author(s) only and do not necessarily reflect those of the European Union or European Innovation Council and SMEs Executive Agency (EISMEA). Neither the European Union nor the granting authority can be held responsible for them. Grant Agreement No: 101099491.}}
\author{\IEEEauthorblockN{Stefano Savazzi\textit{$^{1}$}, Federica Fieramosca\textit{$^{2}$},
Sanaz Kianoush\textit{$^{1}$}, Vittorio Rampa\textit{$^{1}$}, Michele
D'Amico\textit{$^{2}$}} \IEEEauthorblockA{\textit{$^{1}$}\textit{\emph{ Consiglio Nazionale delle Ricerche}}\emph{,}
\textit{\emph{IEIIT}} institute, Piazza Leonardo da Vinci 32, I-20133,
Milano, Italy.\\
 \textit{$^{2}$}\textit{\emph{ }}DEIB, Politecnico di Milano, Piazza Leonardo
da Vinci 32, I-20133, Milano, Italy\linebreak{}
} }
\maketitle
\begin{abstract}
Electromagnetic (EM) body models predict the impact of human presence and motions on the Radio-Frequency (RF) stray radiation received by wireless devices nearby. These wireless devices may be co-located members of a Wireless Local Area Network (WLAN) or even cellular devices connected with a Wide Area Network (WAN). Despite their accuracy, EM models are time-consuming methods which prevent their adoption in strict real-time computational imaging problems and Bayesian estimation, such as passive localization, RF tomography, and holography. Physics-informed Generative Neural Network (GNN) models have recently attracted a lot of attention thanks to their potential to reproduce a process by incorporating relevant physical laws and constraints. Thus, GNNs can be used to simulate/reconstruct missing samples, or learn physics-informed data distributions. The paper discusses a Variational Auto-Encoder (VAE) technique and its adaptations to incorporate a relevant EM body diffraction method with applications to passive RF sensing and localization/tracking. The proposed EM-informed generative model is verified against classical diffraction-based EM body tools and validated on real RF measurements. Applications are also introduced and discussed.
\end{abstract}


\section{Introduction}

\label{sec:intro}

\IEEEPARstart{P}{assive} radio sensing employs stray ambient radio signals from Radio Frequency (RF) devices to detect, locate, and track people that do not need to wear any electronic device, namely device-free~\cite{youssef-2007} -\cite{savazzi-2016}. In line with the \emph{Communication while Sensing} paradigm~\cite{savazzi-2016}, these methods provide seamless detection capabilities, while performing radio communications. In fact, radio signals encode a view of all moving/fixed objects traversed during the signal propagation, and several data analytic methods, such as Bayesian~\cite{kat} and machine learning approaches~\cite{palip}, can be usefully employed to decode this information, typically by large-scale processing of radio signals exchanged by the wireless devices. 

Almost all emerging approaches proposed for solving the radio sensing problem require an approximated knowledge of a physical-informed (prior) model to evaluate the effects of human subjects on radio propagation. The perturbative effects of RF signals induced by the presence or movements of human bodies can be interpreted using Electro-Magnetic (EM) propagation theory considerations~\cite{krupka-1968}. These methods have paved the way to several physical and statistical models for passive radio sensing, which exploit full wave approaches, ray tracing, moving point scattering~\cite{scatt}, and diffraction theory~\cite{koutatis-2010}-\cite{rampa-2022a}. The body-induced perturbations that impair the radio channel can be thus collected, measured, and processed using physical-informed models to estimate location and track target information. A general-purpose EM tool for the prediction of body-induced effects on propagation is still under evaluation~\cite{hamilton-2014}. While simplified or approximated EM models such as path-loss methods~\cite{rodrigues-2021} are too simplistic to capture the complexity of the EM environment, current EM models are too complicated or time-consuming to be of practical use for real-time sensing scenarios~\cite{eleryan-2011}, although usable for off-line applications e.g. during pre-deployment assessment~\cite{kianoush-2016}.

\begin{figure}
\centering \includegraphics[scale=0.41]{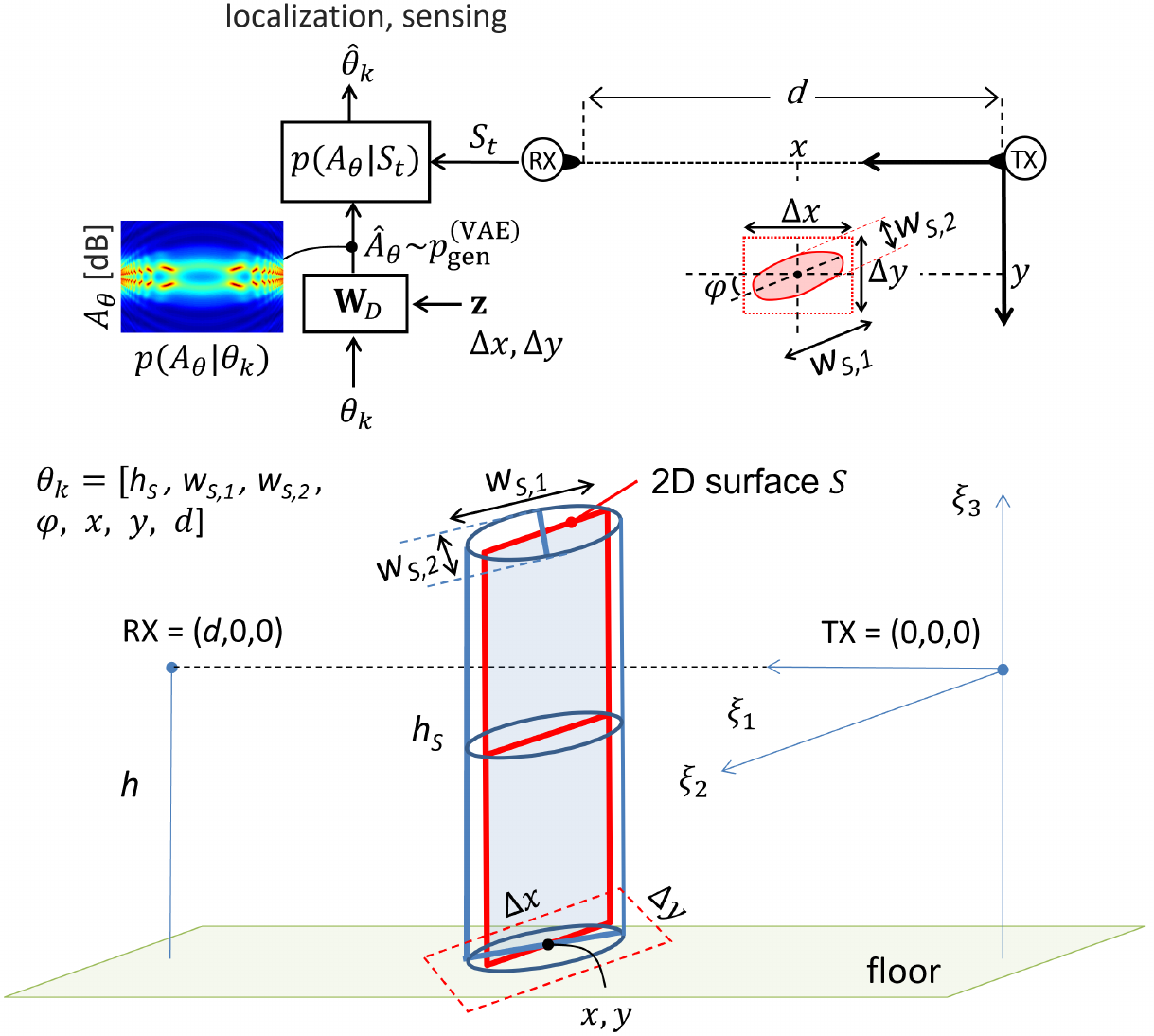} \protect\caption{\label{intro} From top to bottom: the generative approach; the EM model link geometry including the 2D sheet-like obstacle and the RX/TX antennas.}
\end{figure}

Physics-informed generative modelling~\cite{generative_mag} is an emerging field in several application contexts ranging from Bayesian estimation, computational imaging and inverse problems \cite{tomog, channelmodel}. Generative models are typically useful for reproducing the physics of a given phenomenon by generating observations drawn from any prior distribution which reflects the complex underlying physics~\cite{khan} of the environment under study. For the first time, the paper discusses the adoption of an EM-informed generative model inspired by Variational Auto-Encoders (VAE) tools~\cite{bond}. The proposed model is designed to reproduce as close as possible the effects of body movements on RF propagation. 
The proposed generative model is trained by using samples obtained by EM diffraction-based body models~\cite{rampa-2017,eleryan-2011,rampa-2022a}, under different environment configurations and is validated on a real experimental measurements.

The paper is organized as follows: Sect.~\ref{sec:setup} introduces the problem and the proposed setup. Sect.~\ref{sec:Diffraction-models} discusses relevant EM body models for passive RF sensing which make use of diffraction theory considerations and target Bayesian prior distribution modelling. Sect.~\ref{sec:EM-informed-generative-models} describes a generative technique inspired by conditional VAE tools. Sect.~\ref{sec:Model-validation} discusses the implemented generative model structure and verifies its effectiveness in reproducing the EM model diffraction effects considering a passive localization application and real data. Some conclusions are drawn in Sect.~\ref{sec:conclusions}.

\section{Problem setup and motivations\label{sec:setup}}

Radio sensing targets an inverse problem where the goal is to extract the effects ($A_{\theta}$) of the human subject(s) on propagation, using noisy measurements $S_{t}$ of the RF radiation observed at time $t$. The subject(s) are characterized by an unknown state $\mathbf{\theta}$, such as location, size, orientation~\cite{kat,rampa-2017,rampa-2022a}, which we want to recover from the underlying reconstructed data $A_{\theta}$. Under Bayesian formulation, the objective is to maximize the a posterior distribution $p(A_{\theta}|S_{t})$ 
\begin{equation}
p(A_{\theta}|S_{t})=\frac{p(S_{t}|A_{\theta})\cdot p(A_{\theta})}{p(S_{t})}\label{eq:post}
\end{equation}
of the (unknown) data $A_{\theta}$, given the measurements $S_{t}$. Measurements are hardware-specific: they can be in the form of received power, Received Signal Strength (RSS), or base-band Channel State Information (CSI)~\cite{savazzi-2016}. Observations $S_{t}$ are perturbed by the body movements according to a prior-distribution, $p(A_{\theta})$, which models the expected effects of the body (also referred to as the target) in state $\mathbf{\theta}$ as the result of the propagation of the reflected, scattered, and diffracted EM waves. Maximum A-Posteriori (MAP) solution to the inverse problem (\ref{eq:post}) allows to extract the most likely effects $A_{\theta}$, from which it is possible to recover the subject state and any \textit{feature} ($\mathbf{\vartheta}_{k} \in \mathbf{\theta}$) of interest, e.g. the body position, orientation, or size.

The Bayesian approach (\ref{eq:post}) for solving the radio sensing problem requires the knowledge of the model of the RF measurements, or the likelihood function $p(S_{t}|A_{\theta})$, and the prior distribution $p(A_{\theta})$, namely a model to interpret the body-induced EM effects as a function of the features $\mathbf{\theta}$. While the likelihood term depends on the data collection process and on the impairments induced by multipath fading as well, the prior distribution $p(A_{\theta})$ is usually hard to model as it often requires full wave EM approaches, or approximated solutions, which are in many cases too time-consuming to be of practical use for real-time sensing scenarios~\cite{eleryan-2011}.

The paper proposes the use of an EM-informed generative model that is designed to learn the prior distribution of $A_{\theta}$ and reproduce the effects of EM diffraction-based body models~\cite{rampa-2017,rampa-2022a}, under different configurations of the target(s), namely position, size and height, and the environment. The off-line training of the proposed generative model is designed to match the model distribution $p(A_{\theta})$ using (few) samples obtained from an EM body model which is based on the scalar diffraction theory. 

Accurate learning of the prior $p(A_{\theta})$ allows the generative model to reproduce the expected effects of the body on the RF signals under target or link configurations which might be unseen during the training phase, or rather impossible to reproduce through EM simulations. Generative modeling is also well-suited for real-time target tracking implementations as it does not need an ad-hoc generation of EM model samples, that requires intensive computations depending on the target size, number, and the environment (walls, floor, ceiling, and other obstacles). 

\section{Body-induced diffraction effects and Bayesian prior modelling \label{sec:Diffraction-models}}

In this section, we first review and discuss the diffraction-based EM body model~\cite{rampa-2017} used to reproduce the effects of body movements $A_{\theta}$ on propagation, and its applications to Bayesian prior formulation in (\ref{eq:post}). We consider here a single target, but extension to multi-target scenarios can be easily inferred according to~\cite{rampa-2022a}. Also, we will always assume that the monitored target is in the Fraunhofer's region of both transmitting (TX) and receiving (RX) antennas. 

\subsection{EM body models based on diffraction}

As depicted in Fig.~\ref{intro}, we assume that the length of the radio link is given by $d$ while $h$ is its height from the floor. Here, the effects of floor, walls, ceiling or other obstacles are not considered. However, with some effort, these obstacles can be included, as shown in~\cite{fieramosca-2023}. The scalar diffraction theory assumes that the 3-D shape of the human body is modeled as a $2$-D rectangular absorbing sheet $S$~\cite{rampa-2017} with height $h_{S}$ and traversal size that changes according to a 3D cylinder view, with max. and min. traversal sizes $w_{S,1},w_{S,2}$, respectively. The target has nominal position coordinates $\mathbf{p}=[x,y]$, w.r.t. the TX position, which is defined by the projection of its barycenter on the horizontal plane that includes the Line-of-Sight (LoS). The 2-D target might be also rotated of an angle $\varphi$ with respect to the LoS direction. All body features are collected into the vector $\mathbf{\theta}:=\left\{ \mathbf{p},\varphi,h_{S},w_{S,1},w_{S,2}\right\} $ that describes the subject state where time is omitted to simplify the reasoning.
A distribution of Huygens' sources of elementary area $dS$ is located on the absorbing sheet $S$, so that the electric field $E_{\theta}$ at the receiver~\cite{rampa-2017} is obtained by subtracting the contribution of the obstructed Huygens' sources from the electric field $E_{0}$ of the free-space scenario (with no target in the link area):
\begin{equation}
E_{\theta}=E_{0}-\int_{S}dE\label{eq:int}
\end{equation}
According to~\cite{rampa-2017}, (\ref{eq:int}) can be rewritten as:
\begin{equation}
\frac{E_{\theta}}{E_{0}}=1-j\,\frac{d}{\lambda}\,\int_{S}\frac{1}{r_{1}\,r_{2}}\,\exp\left\{ -j\frac{2\pi}{\lambda}\left(r_{1}+r_{2}-d\right)\right\} d\xi_{2}\,d\xi_{3},\label{eq:dE_full_compact-1-1-1}
\end{equation}
where $\lambda$ is the wavelength. Notice that each elementary source $dS=d\xi_{2}\,d\xi_{3}$ has distance $r_{1}$ and $r_{2}$ from the TX and the RX, respectively which depend on the relative coordinates $\mathbf{p}$.

The RSS measurement $S_{t}$, defined at the generic frequency $f$ (and time $t$, here omitted again for clarity) is obtained as:

\begin{equation}
\begin{aligned}S_{t}={} & \left\{ \begin{array}{ll}
P_{0}+w_{0} & \;\textrm{free-space only}\\
P_{0}-A_{\theta}+w_{T} & \;\textrm{with target }S
\end{array}\right.,\\
A_{\theta}={} & -10\,\log_{10}\left|\frac{E_{\theta}}{E_{0}}\right|^{2},
\end{aligned}
\label{eq:received_power-1}
\end{equation}
where $A_{\theta}$ is the excess attenuation due to the presence of $S$, $P_{0}$ is the free-space power that depends on the link geometry and $w_{0}$ and $w_{T}$ model the log-normal multipath fading and the other disturbances. Noise terms are Gaussian distributed where $w_{0}\sim\mathcal{N\mathrm{\left(0,\mathit{\sigma_{\textrm{0}}^{2}}\right)}}$, with variance $\mathrm{\mathit{\sigma_{\textrm{0}}^{2}}}$, refers to the free-space case only and $w_{T}\sim\mathcal{N\mathrm{\left(\mathit{\mu_{T},\sigma_{T}^{2}}\right)}}$, with mean $\mu_{T}=\Delta h_{T}$ and variance $\mathrm{\mathit{\sigma_{T}^{2}}}=\mathit{\sigma_{\textrm{0}}^{2}+\Delta\sigma_{T}^{2}}$, to the case with the target. $\Delta h_{T}$ and $\Delta\sigma_{T}^{2}\geq0$ are the residual stochastic fading terms that depend on the specific scenario as shown in \cite{rampa-2017}. 

\subsection{Bayesian prior distribution}

Models (\ref{eq:dE_full_compact-1-1-1}) and (\ref{eq:received_power-1}) produce an observation of the excess attenuation $A_{\theta}$ as caused by a body under the state $\theta$. Imperfect knowledge of the environment, small body movements, or changing configurations, make body features hard to obtain with an acceptable level of accuracy. In addition, in radio sensing applications, we are often interested to recover a subset of body features, e.g. the subject locations $\mathbf{p}$ or the obstruction size ($h_{S},w_{S,1},w_{S,2}$), while leaving the others partially (or fully) unknown. Here, we resort to a statistical approach where $A_{\theta}$ is sampled from a distribution $p(A_{\theta})$ defined as:
\begin{equation}
p(A_{\theta}|\theta_{k})=A_{\theta\sim p(\theta|\theta_{k})}.\label{eq:samp}
\end{equation}
The excess attenuation $A_{\theta}$ is thus obtained for random instances of $\theta$ which follow the probability function $p(\theta|\theta_{k})$. The function $p(\theta|\theta_{k})$ models the uncertainty with respect to the \emph{nominal features} $\theta_{k}$. For example, in passive localization, the complex structure of the human body and the difficulties to measure its true position makes the footprint of the nominal (i.e., measured) location $\theta_{k}=\mathbf{p}_{k}$ of the target subject to an error in the $\xi_{1}$ and $\xi_{2}$ directions in the order of $5\div10$ cm~\cite{rampa-2022a}. The prior distribution $p(A_{\theta}|\theta_{k})$ is set to capture such small, but not measurable, movements, i.e., with $\theta\sim p(\theta|\theta_{k})$ as uniformly distributed in an elementary area of size $\triangle x,\triangle y\,=\,10\div20$ cm. Similar reasoning holds for other body features as well. Further examples are given in Sect.~\ref{sec:Model-validation}.

\begin{figure}
\centering \includegraphics[scale=0.37]{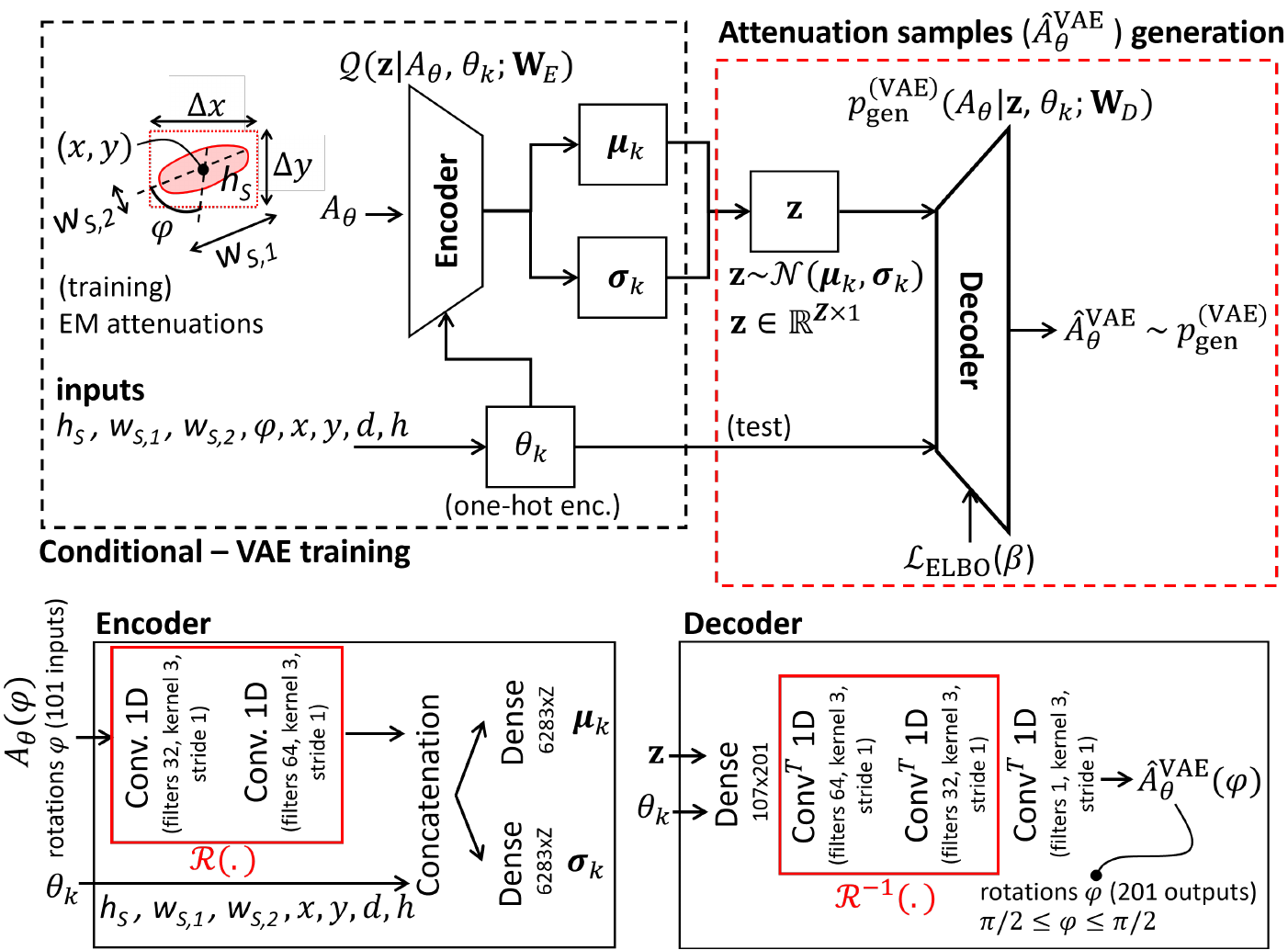} \protect\caption{\label{vae_gan} Conditional VAE (C-VAE) architecture for generating body induced excess attenuation samples. Bottom: encoder and decoder neural network structures. Dense, Conv and Conv$^{T}$ refers to fully connected, convolution and deconvolution layer \cite{deconv} operators, respectively.}
\end{figure}

\section{EM-informed Conditional Variational AutoEncoder (C-VAE)\label{sec:EM-informed-generative-models}}

The generative model considered here is able to reproduce the RSS excess attenuation values $A_{\theta}$ (\ref{eq:received_power-1}) as if they are sampled from the \emph{ground-truth} prior distribution $p(A_{\theta}|\theta_{k})$ in (\ref{eq:samp}). A conditional-VAE (C-VAE) network \cite{bond} is proposed and adapted by conditioning the generative process on the input features $\theta_{k}$, so to generate samples close to the conditional prior $p(A_{\theta}|\theta_{k})$. As shown in Fig.~\ref{vae_gan}, the process is implemented by a \emph{decoder} (VAE) parameterized by the neural network (NN) parameters $\mathbf{W}_{D}$ and an \emph{encoder} $\mathcal{Q} (\mathbf{z}|A_{\theta},\theta_{k};\mathbf{W}_{E})$ parameterized by the NN parameters $\mathbf{W}_{E}$. The decoder maps an input latent space $\mathbf{z}\sim p_{\mathcal{Z}}(\mathbf{z})$ of size $Z$ ($\mathbf{z}\in\mathbb{R}^{Z\times1}$), into an output space $\widehat{A}_{\theta}\sim p_{\mathrm{gen}}(A_{\theta}|\theta_{k})$ where the generated distribution $p_{\mathrm{gen}}$ is set to reproduce the targeted EM model, namely $p_{\mathrm{gen}}(A_{\theta}|\theta_{k})\cong p(A_{\theta}|\theta_{k})$, or equivalently $\widehat{A}_{\theta}\cong A_{\theta}$ for all inputs $\theta_{k}$ of interest. The \emph{encoder} learns the latent space $p_{\mathcal{Z}}(\mathbf{z}|\theta_{k})=\mathcal{N}(\mathbf{\boldsymbol{\mu}_{\mathit{k}}},\boldsymbol{\sigma}_{k}^{2})$ which is a multivariate Gaussian distribution with mean $\mathbf{\boldsymbol{\mu}_{\mathit{k}}}$ and standard deviation $\boldsymbol{\sigma}_{k}$ 
(other choices are possible but they are not investigated here). The encoder uses the training samples of body-induced excess attenuation values $A_{\theta}$ obtained from the EM model (\ref{eq:received_power-1}) and the corresponding features $\theta_{k}$. Encoder and decoder models are shown in Fig.~\ref{vae_gan} at bottom. Pre-trained neural network parameters are available on-line \cite{github} together with example codes for testing. 

In this paper, we limit our focus on simple body features $\theta_{k}=\left[\mathbf{p}_{k},\varphi_{k},h_{S},w_{S,1},w_{S,2}\right]$ that include specific body locations $\mathbf{p}_{k}$, relative orientations $-\pi/2\leq\varphi_{k}\leq\pi/2$, and different sizes $h_{S},w_{S,1},w_{S,2}$ of the target. 
Although more complex approaches are possible, notice that even with such a simple feature set, the conditional prior $p(A_{\theta}|\theta_{k})$ obtained as in (\ref{eq:samp}) with uncertainties $p(\theta|\theta_{k})$ is complex enough to make a full EM simulation unfeasible, thus motivating the use of generative methods. Besides, the decoder can reproduce body-induced attenuation samples conditioned on the features $\theta_{k}$ that are unknown at training time, i.e., due to new subject locations, orientations or target sizes. 

The C-VAE decoder produces a distribution $\widehat{A}_{\theta}^{\mathrm{VAE}}\sim p_{\mathrm{gen}}^{\mathrm{VAE}}(A_{\theta}|\theta_{k})$
\begin{equation}
p_{\mathrm{gen}}^{\mathrm{VAE}}(A_{\theta}|\theta_{k})=\int_{\mathcal{Z}}p_{\mathrm{gen}}^{\mathrm{VAE}}(A_{\theta}|\mathbf{z},\theta_{k};\mathbf{W}_{D})\,p_{\mathcal{Z}}(\mathbf{z}|\theta_{k})\,d\mathbf{z},\label{eq:vaegen}
\end{equation}
which is the marginalization of the conditional probability $p_{\mathrm{gen}}^{\mathrm{VAE}}(A_{\theta}|\mathbf{z},\theta_{k};\mathbf{W}_{D})$ function of the NN parameters $\mathbf{W}_{D}$. The goal is to maximize the likelihood bound called Evidence Lower BOund (ELBO) $\mathcal{L}_{\mathrm{ELBO}}$ described in \cite{vae-1}: omitting dependency on parameters $\mathbf{W}_{E}$
and $\mathbf{W}_{D}$, it is
\begin{equation}
\mathcal{L}_{\mathrm{ELBO}}=\ell_{k}-\beta\cdot\mathrm{D_{KL}}\left[\mathcal{Q}(\mathbf{z}|A_{\theta},\theta_{k})\Vert p_{\mathcal{Z}}(\mathbf{z}|\theta_{k})\right].\label{eq:elbo}
\end{equation}
The first term $\ell_{k}=\mathbb{E}_{\mathbf{z}\sim\mathcal{Q}(\cdot|)}\mathrm{log}\left[p_{\mathrm{gen}}^{\mathrm{VAE}}(A_{\theta}|\mathbf{z},\theta_{k})\right]$ is the log-likelihood, while the second one it the Kullback-Leibler (KL) divergence $\mathrm{D_{KL}}$ \cite{kull} between the encoder output and the input latent space. The ELBO metric (\ref{eq:elbo}) is then averaged over the input training features $\theta_{k}$. 

The maximization of the likelihood $\ell_{k}$ makes the generated samples $\widehat{A}_{\theta}^{\mathrm{VAE}}$ more correlated to the latent variables $\mathbf{z}$, which typically cause the model to be more deterministic. On the other hand, the weight term $\beta>0$ can be used to increase the contribution of the KL divergence between the posterior and the prior to the total ELBO and thus increase the randomness of generated samples. In Sect.~\ref{sec:Model-validation}, we will show that the weight term $\beta$ can be optimized targeting passive localization applications where the RSS measurements $S_{t}$ are affected by noise and multipath interference. 

\begin{figure}
\centering \includegraphics[scale=0.55]{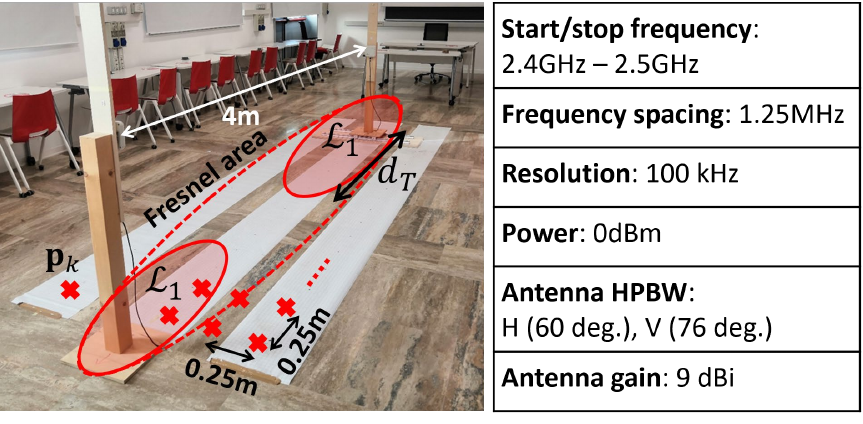} \protect\caption{\label{scenario} Measurement setup, explored target positions and Fresnel's area.}
\end{figure}
\begin{figure}
\centering \includegraphics[scale=0.55]{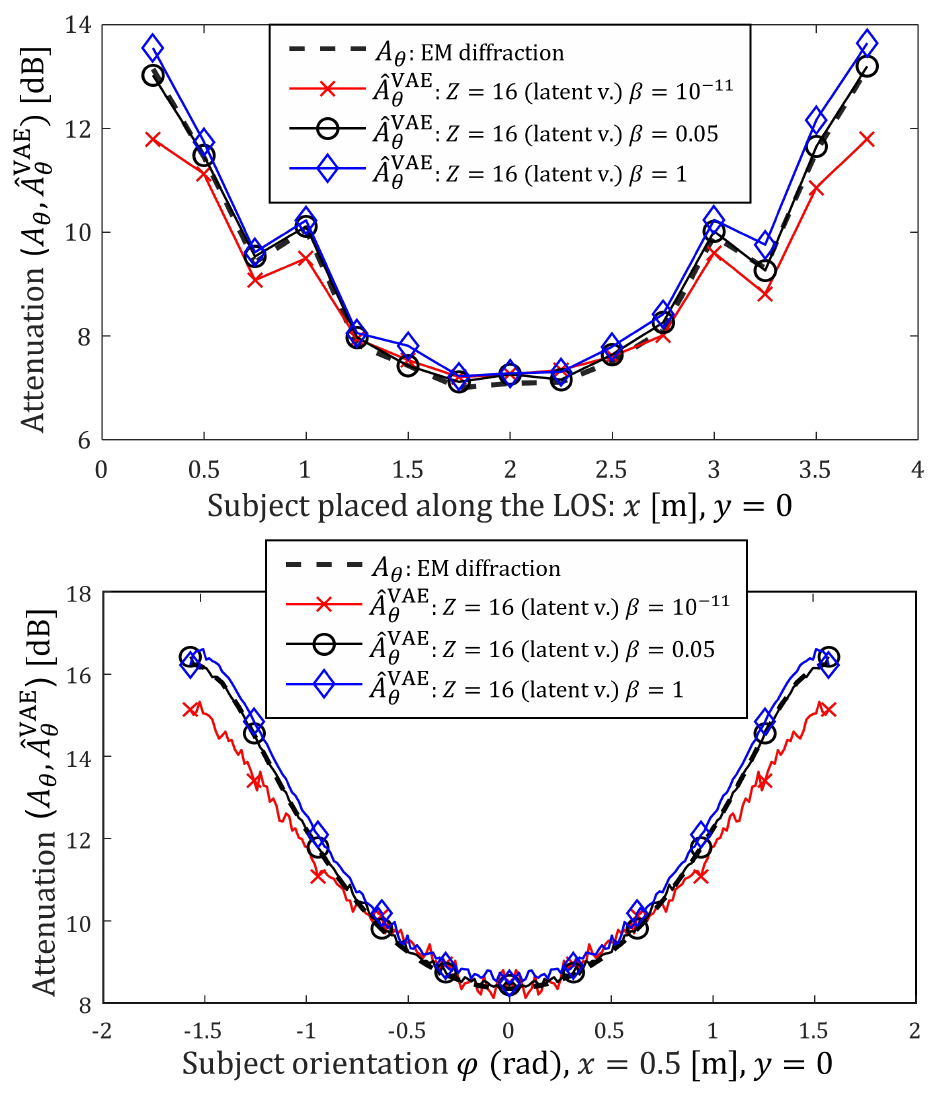} \protect\caption{\label{vae_gan-1} C-VAE generation of EM body model ($h_{S}=1.80\mathrm{m},w_{S,1}=0.55 \mathrm{m},w_{S,2}=0.25\mathrm{m}$), $Z=16$ latent samples, and varying $\beta$. (a) The subject is moving along the LOS ($0.25\mathrm{m}\protect\leq x\protect\leq3.75\mathrm{m}$, $y=0$). The EM body excess attenuation values $A_{\mathbf{\theta}}$ for model training are obtained by averaging over random target orientations $-\pi/2\protect\leq\varphi\protect\leq\pi/2$ and random movements in an elementary area of size $\triangle x=\triangle y=0.1\mathrm{m}$. (b) The target is in position $x=0.5\mathrm{m}$, $y=0$ and changing orientation $\varphi$ while performing small movements in the same elementary area.}
\end{figure}
\begin{figure}
\centering \includegraphics[scale=0.6]{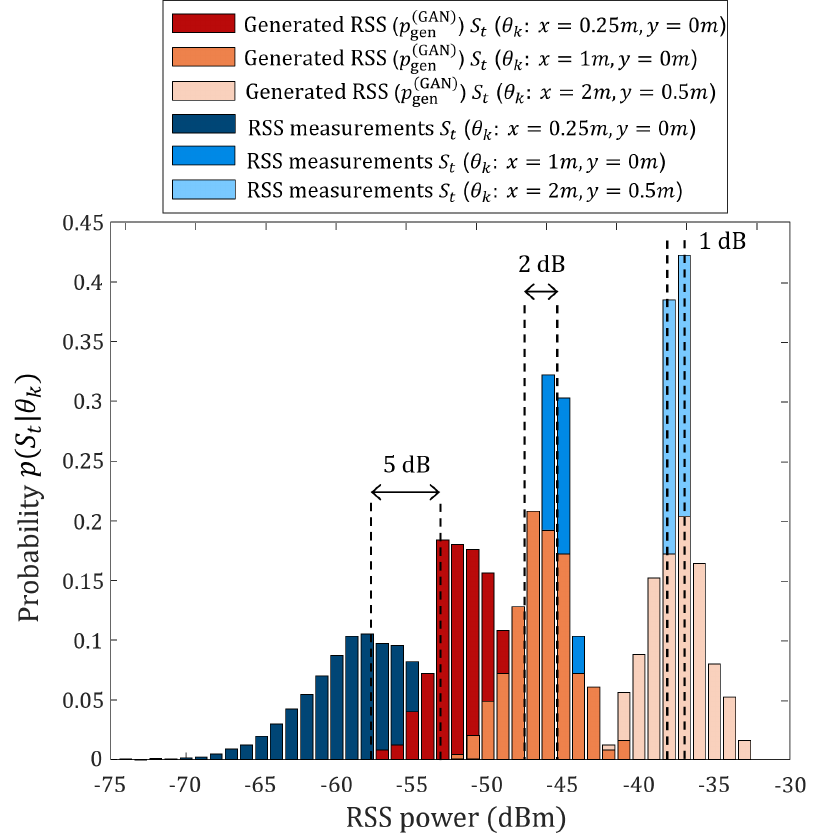} \protect\caption{\label{vae_gan-1-1} C-VAE generation of RSS samples (same target size as Fig.~\ref{scenario} and $\varphi=0$) with $\beta=1$ (red bars) compared with true RSS measurements (blue bars) obtained at 2.4GHz. Generated RSSs are obtained by marginalization (\ref{eq:marg}) with  $p_{\mathrm{gen}}^{\mathrm{VAE}}(A_{\theta}|\theta_{k})$ in (\ref{eq:vaegen}) and likelihood $p(S_{t}|A_{\theta})$ in (\ref{eq:received_power-1}) with $\mu_{T}=2$dB, $\mathrm{\mathit{\sigma_{T}}=2}$dB and $\mathit{\sigma}_{0}=1$dB. The subject is standing while performing small movements around $3$ nominal positions $\mathbf{p}_{k}=(x,y)$ detailed above.}
\end{figure}

\section{Model accuracy and validation\label{sec:Model-validation}}

The generative modelling approach has been validated with measurements taken in a hall of size $6.1\mathrm{m}$ $\times$ $14.4\mathrm{m}$ as shown in Fig.~\ref{scenario}. TX and RX nodes are spaced $d=4\mathrm{m}$ apart, while the LOS is parallel to the lateral walls and horizontally placed at $h=0.99\mathrm{m}$ from the floor. Both TX and RX are equipped with directional antennas with parameters summarized in the Table of Fig. \ref{scenario}. The human target (one of the authors who volunteered) is modelled as a rectangular $2$-D sheet with height $h_{S}=1.80\mathrm{m}$ and traversal max. and min. sizes equal to $w_{S,1}=0.55\mathrm{m}$ and $w_{S,2}=0.25\mathrm{m}$, respectively. The received power $S_{t}$ is measured using a tracking generator enabled spectrum analyzer \cite{tek} in the $2.4\div2.5$GHz band, over 81 frequencies with $1.25$MHz spacing; for each frequency, 500 consecutive time samples are acquired in 1 minute (120ms sampling time). The target is located in $K=75$ marked positions $\mathbf{p}_{k}$, $k=1,...,K$, of the grid points of Fig.~\ref{scenario}. The spacing between marked positions is $0.25\mathrm{m}$ along and across the LOS. 

In the proposed C-VAE implementation of Fig.~\ref{vae_gan}, the decoder consists of approx. $20\mathrm{K}$ trainable parameters and has size of $1$MB which is enough for real time sample generation. The encoder has 420K trainable parameters and size $5$MB. Although out of the scope of the current paper, accurate model pruning is desirable to minimize the memory footprint on resource-constrained devices~\cite{savazzi-2016}. In what follows, the C-VAE model accuracy is first addressed against EM body attenuation samples obtained from both the diffraction model and the RSS measurements. Next, we highlight an application in passive radio localization.

\subsection{Generative model assessment}

Fig.~\ref{vae_gan-1} shows an example of C-VAE generation of the diffraction model using $Z=16$ latent samples and different ELBO weights $\beta$. In Fig.~\ref{vae_gan-1}(a), the C-VAE model is used to reproduce the attenuation samples $\widehat{A}_{\theta}^{\mathrm{VAE}}$ corresponding to a subject that is moving along the LOS. Results are compared with EM body attenuation values $A_{\mathbf{\theta}}$ obtained from (\ref{eq:received_power-1}) in the corresponding positions (dashed line). Samples are averaged over random target orientations $-\pi/2\leq\varphi\leq\pi/2$ and movements in an elementary area of size $\triangle x=\Delta y=0.1\mathrm{m}$. In Fig.~\ref{vae_gan-1}(b), the target is now fixed in position $x=0.5\mathrm{m}$, $y=0$ which changing orientation $\varphi$. As previously introduced, the weighting factor $\beta$ in $\mathcal{L}_{\mathrm{ELBO}}$ of (\ref{eq:elbo}) affects the generated samples: looking to both examples, $\beta=0.05$ stands as a good compromise between average reproduction accuracy and minimization of posterior distribution divergence.

Generation times of C-VAE and diffraction model are compared in Tab.~\ref{generation_times}. Time measurements are obtained using a Jetson Nano single-board computer equipped with a quad-core ARM-Cortex-A57 SoC, 4 GB RAM, 128 CUDA cores, and a Maxwell GPU architecture which is representative of a typical resource-constrained IoT device. Note that, for all cases, the C-VAE generation of attenuation samples is about $\times\,60\div100$ times faster than EM model computation, which also depends on the chosen numerical integration configurations (i.e., tiled integration method, and absolute error tolerance), target size, and antenna configuration (omnidirectional vs directional radiation patterns). A well-trained generative model can be therefore used to simulate the desired prior distribution in real-time, with sufficiently high randomness of samples.

In Fig.~\ref{vae_gan-1-1}, we now adopt the C-VAE model to generate RSS samples $\widehat{S}_{t}\sim p(S_{t}|\theta_{k})$, rather than attenuation values. These are taken by marginalization:
\begin{equation}
p(S_{t}|\theta_{k})=\int_{A_{\theta}}p(S_{t}|A_{\theta})\,p_{\mathrm{gen}}^{\mathrm{VAE}}(A_{\theta}|\theta_{k})\,dA_{\theta},\label{eq:marg}
\end{equation}
with $p_{\mathrm{gen}}^{\mathrm{VAE}}(A_{\theta}|\theta_{k})$ in
(\ref{eq:vaegen}) and likelihood $p(S_{t}|A_{\theta})$ in (\ref{eq:received_power-1}) with $\mu_{T}=2$dB, $\mathrm{\mathit{\sigma_{T}}=2}$dB and $\mathit{\sigma}_{0}=1$dB~\cite{rampa-2022a}. The subject is standing while performing small movements around $3$ selected positions $\theta_{k}=\mathbf{p}_{k}$ detailed in the same figure. The figure compares the corresponding probability mass functions obtained from generated samples $\widehat{S}_{t}$ and the RSS values $S_{t}$ according to the setup in Fig.~\ref{scenario}. The model optimized as in Fig.~\ref{vae_gan-1} is able to effectively reproduce the RSS values in real-time. On the other hand, as also observed with diffraction models \cite{rampa-2017}, the C-VAE generation tool seems to under-estimate the observed RSS values for target positions very close to the transmitter ($x=0.25\mathrm{m}$). Although not considered in this paper, an improved (or ad-hoc) training stage is recommended for these locations.

\begin{table}[tp]
\protect\caption{\label{generation_times}EM-informed C-VAE vs diffraction model: body-induced
attenuation generation time analysis.}
\vspace{0.3cm}
\begin{centering}
\begin{tabular}{|c|c|c|}
\hline 
\multicolumn{1}{|c|}{\textbf{Model}} & \multicolumn{1}{c|}{\textbf{Parameters}} & \multicolumn{1}{c|}{\textbf{Average generation }}\tabularnewline
 &  & \textbf{time }{[}s/sample{]}\tabularnewline
\hline 
\multirow{2}{*}{\textbf{C-VAE generation}} & $Z=16$  & \multicolumn{1}{c|}{$3.5\times10^{-5}$ }\tabularnewline
 & $Z=32$  & $5.2\times10^{-5}$\tabularnewline
\hline 
\textbf{EM Diffraction }  & Abs err. tol.=$10^{-3}$  & $5.4\times10^{-3}$ \tabularnewline
(omnidirectional antennas)  & Abs err. tol.=$10^{-6}$  & $3.81\times10^{-2}$ \tabularnewline
\hline 
\textbf{EM Diffraction }  & Abs err. tol.=$10^{-3}$  & $0.64$ \tabularnewline
(directional antennas)  & Abs err. tol.=$10^{-6}$  & $1.56$ \tabularnewline
\hline 
\end{tabular}
\par\end{centering}
\medskip{}
\vspace{-0.6cm}
\end{table}

\subsection{Applications to passive localization}

In the following, we discuss an example of passive localization where
the goal is to detect the target presence and assign the most likely
motion area. With this respect, as depicted in Fig. \ref{scenario},
we define two regions in the space surrounding the link: the first
one contains the positions $\mathbf{p}_{k}\in\mathcal{L}_{0}$ that
fall outside the Fresnel's ellipsoid, the second one, the positions
$\mathbf{p}_{k}\in\mathcal{L}_{1}(d_{T})$ that fall inside the ellipsoid,
while being at a distance of $d_{T}<d$ from the TX (or RX). Using
C-VAE generated samples from $p_{\mathrm{gen}}^{\mathrm{VAE}}$, first,
we estimate the most likely body effects $\widehat{A}(\theta_{k})$
under the assumption that the target is in state $\theta_{k}$ 
\begin{equation}
\widehat{A}(\theta_{k})=\mathrm{max}_{\theta}p(A_{\theta}|S_{t},\theta_{k})=\mathrm{max}_{\theta}p(S_{t}|A_{\theta})p_{\mathrm{gen}}^{\mathrm{VAE}}(A_{\theta}|\theta_{k}).\label{eq:map}
\end{equation}
Next, we recover the target position $k$ by MAP estimation: 
\begin{equation}
\widehat{\mathbf{p}}_{k}=\mathrm{argmax}_{k}\widehat{A}(\theta_{k}=\mathbf{p}_{k}),\label{eq:EST}
\end{equation}
or, equivalently, $\widehat{\mathbf{p}}_{k}=\mathrm{argmax}_{k}\left[\mathrm{max}_{\theta}p(A_{\theta}|S_{t},\theta_{k})\right]$,
and make a decision based on the two hypothesis, namely $\mathcal{H}_{0}$:
$\widehat{\mathbf{p}}_{k}\in\mathcal{L}_{0}$, $\mathcal{H}_{1}$:
$\widehat{\mathbf{p}}_{k}\in\mathcal{L}_{1}$.

Using RSS samples collected from measurements, Tab.~\ref{table_performance_vae} analyzes the detection probability
of a target outside the Fresnel area, namely $p_{\mathcal{L}_{0}}=\mathrm{Pr}[\widehat{\mathbf{p}}_{k}\in\mathcal{L}_{0}|\mathbf{p}_{k}\in\mathcal{L}_{0}]$,
and inside the same area, at distance $d_{T}$ from the TX/RX, $p_{\mathcal{L}_{1}}(d_{T})=\mathrm{Pr}[\widehat{\mathbf{p}}_{k}\in\mathcal{L}_{1}|\mathbf{p}_{k}\in\mathcal{L}_{1}]$.
The purpose is to evaluate how performance, i.e., for varying $d_{T}$,
is affected by the C-VAE model generation, focusing in particular
on the choice of the number of latent variables $Z$ and the weighting
factor $\beta$. According to these tests a target at a distance of $d_T = 1$m from the TX (or RX) can be easily detected by imposing a number of latent variables $Z \geq 16$ with a detection probability of $p_{\mathcal{L}_{0}}(d_{T}=1m) \approxeq 0.7$ and $p_{\mathcal{L}_{1}}(d_{T}=1m)>0.85$ on average for $\beta \geq 0.05$.

\begin{table}[tp]
\protect\caption{\label{table_performance_vae}Detection probability for a target outside
($p_{\mathcal{L}_{0}}$), close to TX/RX $(p_{\mathcal{L}_{1}})$
at varying distance $d_{T}$ for different number of latent variables
$Z$ and weighting factor $\beta$.}
\vspace{0.3cm}

\begin{centering}
\begin{tabular}{|l|l||c|c|c|}
\cline{3-5} 
\multicolumn{1}{c}{} & \multicolumn{1}{c|}{} & \multicolumn{3}{c|} {\centering \textbf{$p_{\mathcal{L}_{0}}(d_{T})$ $/$ $p_{\mathcal{L}_{1}}(d_{T})$} }\tabularnewline
\hline
\multicolumn{2}{|c||}{\textbf{C-VAE parameters}} & \multirow{2}{*}{$0.75$ m} & \multirow{2}{*}{$1$ m}  & \multirow{2}{*}{$1.25$ m}  \tabularnewline
\multicolumn{2}{|l||}{} &  &  &  \tabularnewline
\hline 
$Z=8$  & $\beta\simeq0$  & $0.72$$/$$0.55$  & $0.68$$/$$0.85$ & $0.72$$/$$0.55$  \tabularnewline
$Z=8$  & $\beta=0.05$  &  $0.69$$/$$0.73$  & $0.69$$/$$0.86$  & $0.81$$/$$0.99$\tabularnewline
$Z=8$  & $\beta=1$  & $0.70$$/$$0.98$ & $0.71$$/$$0.86$  & $0.70$$/$$0.98$  \tabularnewline
\hline 
$Z=16$  & $\beta\simeq0$ & $0.69$$/$$0.99$  & $0.71$$/$$0.98$  & $0.70$$/$$0.99$  \tabularnewline
$Z=16$  & $\beta=0.05$  & $0.70$$/$$0.73$ & $0.69$$/$$0.86$  & $0.70$$/$$0.73$  \tabularnewline
$Z=16$  & $\beta=1$  & $0.70$$/$$0.73$  & $0.69$$/$$0.86$  & $0.70$$/$$0.73$ \tabularnewline
\hline 
$Z=32$  & $\beta\simeq0$  &  $0.70$$/$$0.67$  & $0.68$$/$$0.99$  & $0.69$$/$$0.67$ \tabularnewline
$Z=32$  & $\beta=0.05$  & $0.71$$/$$0.68$  & $0.70$$/$$0.86$  & $0.71$$/$$0.68$  \tabularnewline
$Z=32$  & $\beta=1$  & $0.71$$/$$0.73$  & $0.70$$/$$0.88$  & $0.70$$/$$0.73$  \tabularnewline
\hline 
$Z=48$  & $\beta\simeq0$  &  $0.68$$/$$0.78$ & $0.69$$/$$0.70$ & $0.68$$/$$0.78$  \tabularnewline
$Z=48$  & $\beta=0.05$  & $0.69$$/$$0.98$ & $0.67$$/$$0.86$  & $0.67$$/$$0.98$  \tabularnewline
$Z=48$  & $\beta=1$  & $0.71$$/$$0.67$ & $0.69$$/$$0.83$ & $0.71$$/$$0.67$  \tabularnewline
\hline 
\end{tabular}
\par\end{centering}
\medskip{}
 \vspace{-0.6cm}
 
\end{table}

\section{Conclusions and future activities}

\label{sec:conclusions}

The paper proposed the use of an EM-informed generative model tailored for radio sensing applications. The model is designed to learn the Bayesian prior distribution of body induced diffraction effects, so to reproduce the effects of EM diffraction under different body configurations and propagation settings. We considered a Conditional Variational AutoEncoder (C-VAE) tool where the generated distribution is set to generate samples of the targeted EM model through latent variable encoding/decoding neural network operations. The C-VAE model is optimized to improve the process of sampling from the Bayesian prior distribution, namely to achieve an effective compromise between the average reproduction accuracy and the randomness of generated attenuation values.

The generative modelling method has been validated with RSS measurements taken in an indoor site. With respect to complexity, model size and attenuation sample generation time, the proposed C-VAE approach is well-suited for real-time target tracking implementations as it does not require intensive or ad-hoc EM computations. Beside data augmentation for Bayesian prior modelling, the generation approach is also able to predict body-induced attenuation values under propagation and/or target configurations that are not seen during training. Preliminary examples tailored for passive localization  reveal the possibility of optimizing the generation process to improve performance. This aspect will be evaluated in more details in future works.

\end{document}